\documentstyle[12pt]{article}
\newcommand{\beq}[1]{\begin{equation}\label{#1}}
\newcommand{\eeq}{\end{equation}}
\newcommand{\bear}[1]{\begin{eqnarray}\label{#1}}
\newcommand{\ear}{\end{eqnarray}}
\newcommand{\nn}{\nonumber}

\newcommand{\rf}[1]{(\ref{#1})}
\newcommand{\nl}{ {\hfill \break} }
\newcommand{\np}{ {\newpage } }

\newcommand{\iso}{ {\cong } }

\newcommand{\Iff}{ {\Leftrightarrow } }

\newcommand{\imp}{\ {\Rightarrow }\ }

\newcommand{\partl}{ {\partial } }
\newcommand{\cl}{ { \mbox{\rm cl} } }

\newcommand{\N}{ \mbox{\rm I$\!$N} }
\newcommand{\R}{ \mbox{\rm I$\!$R} }
\newcommand{\Diff}{ \mbox{\rm Diff} }
\newcommand{\Out}{ \mbox{\rm Out} }
\newcommand{\Inn}{ \mbox{\rm Inn} }
\newcommand{\Aut}{ \mbox{\rm Aut} }

\newcommand{\tr}{ \mbox{\rm tr} }

\def\Journal#1#2#3#4{{#1} {\bf #2}, #3 (#4)}

\def\CMP{\em Commun. Math. Phys.}
\def\CQG{\em Class. Quant. Grav.}
\def\IJTP{\em Int. J. Theor. Phys.}
\def\JMP{\em J. Math. Phys.}

\def\RMP{\em Rev. Math. Phys.}

\def\be{\begin{equation}}
\def\ee{\end{equation}}
\def\bea{\begin{eqnarray}}
\def\eea{\end{eqnarray}}

\begin{document}
\title{\bf\large THE ROLE OF DILATIONS IN DIFFEOMORPHISM\\ 
COVARIANT ALGEBRAIC QUANTUM FIELD THEORY
\footnote{Lecture given at MG8, Jerusalem 1997}}

\author
{M. RAINER\\ 
Mathematische Physik I, Mathematisches Institut\\
Universit\"at Potsdam, PF 601553, D-14415 Potsdam, Germany
}
\date{July 1997}

\maketitle
\vspace*{-8.5cm}
\centerline{\mbox{\hspace*{10cm} Preprint Univ. Potsdam }}
\centerline{\mbox{\hspace*{10cm} July 1997
}}
\centerline{\mbox{\hspace*{10cm} gr-qc/9710081}}
\vspace*{9.0cm}
\centerline{\bf Abstract}{
The quantum analogue of 
general relativistic geometry should be implementable on smooth manifolds 
without an a priori metric structure, the kinematical covariance
group acting by diffeomorphisms.
\nl
Here I approach quantum gravity (QG) in the view of constructive, 
algebraic quantum field theory (QFT). Comparing QG with usual QFT, 
the algebraic approach clarifies analogies 
and peculiarities.
As usual, an isotonic net of $*$-algebras is taken to encode
the quantum field operators. For QG, the kinematical covariance group acts
via diffeomorphisms on the open sets of the manifold, 
and via algebraic isomorphisms on the algebras.
In general, the algebra of observables is covariant 
only under a (dynamical) subgroup 
of the general diffeomorphism group. 
\nl
After an algebraic implementation of
the dynamical subgroup of dilations,
small and large scale cutoffs  may be introduced 
algebraically.  
So the usual a priori conflict of cutoffs with general covariance
is avoided. Even more, these cutoffs provide a natural local cobordism for 
topological quantum field theory.
\nl
A new commutant duality
between the minimal and maximal algebra
allows to extract the modular structure from the net of algebras.
The outer modular isomorphisms are then again related to dilations,
which (under certain conditions) may provide a notion of time.}
\np
\section{Introduction} 
The following investigations might be seen as an attempt to understand
some aspects of quantum field theory (QFT) on differentiable manifolds. 
This is indeed a very promising approach 
to quantum general relativity\cite{Ro}. 

The {observation procedures} 
represent the abstract kinematical framework
for {possible} preparations of measurements, while the {observables}
encode the kinds of questions one can ask from the physical system.
The covariance group of the observation procedures reflects a 
general (a priori) redundancy of their mathematical implementation. 
The more sophisticated the structure of the
observation procedures, the smaller the covariance group will be in general.
In a concrete observation the kinematical covariance will be broken.
So, e.g., in a treelike network of string world tubes 
a concrete local observation 
requires the explicit selection of one of many a priori equivalent vertices,
whence it breaks the covariance which holds for the network of vertices
as a whole\cite{MR}.
However, irrespectively of the loss of 
covariance in a concrete observation, the action
of the covariance group may still be well defined on the observation 
procedures.
In any case, the loss of covariance in a concrete observation is related
to a specific structure of the state of the physical system.

Previous attempts\cite{FrHa,Sal}  
to implement kinematical general covariance
and its dynamical breakdown 
in the spirit of an algebraic, constructive approach\cite{HaKa}  
to quantum (field) theory 
have been continued recently\cite{Sal1,Ra1,RaSa}.
The principle of locality
is kept by demanding that, observation procedures 
correspond to {possible} preparations of 
localized measurements in bounded regions. 
Note that 
here is no a priori notion of 
neither a metric, nor a time, nor even a causal structure. 
Then, on different regions there will be no a priori 
causal relations between observables. 

For a net of subalgebras of a Weyl algebra, it is indeed possible\cite{Ba1} 
to work with a flexible notion of causality  
rather than with a rigidly given one.
In principle it might be possible to construct the net together with
its underlying manifold from a partial order via inclusion
of the algebras themselves only\cite{Ba2}.

Nevertheless, below we start just from a net of $*$-algebras on a 
differentiable manifold.
On this net, a physical state induces dynamical relations,
whence the algebra of observables is covariant just under a
a certain subgroup 
of the general diffeomorphism group.
This subgroup describes a covariance related to the 
(dynamically relevant) observables.
The present examinations emphasize on
the dynamical subgroup of dilations.

For QFT on Minkowski space (see Sect. 3 below)
the usual causality condition
can be sharpened to {Haag duality}\cite{Ha},
which there in particular opens the door for the powerful 
DHR-analysis\cite{DHR},
Haag duality is the first example of an algebraic commutant duality
on a net of von Neumann algebras.
Recently\cite{Sal1,Ra1,RaSa}, 
attempts were made to find an appropriate 
modification of Haag duality that can be applied 
in consistency with general covariance on
an arbitrarily curved background.
In Sect. 4 below, the idea of a certain  algebraic commutant duality
between very large scales and very small scales is exploited,
in order to obtain a modular group of algebraic
isomorphisms on a net of von Neumann algebras.   

In \cite{Sal1} such an idea was implemented
for the asymptotical large scale limit $n\to\infty$ 
of discretely parametrized domains ${\cal O}_n$, $n\in \N$.

Unlike there, in \cite{Ra1,RaSa} a von Neumann commutant duality 
${\cal R}({\cal O}^x_{s_{\min}})={\cal R}'({\cal O}^x_{s_{\max}})$ 
was introduced between some minimal and some maximal algebra 
from a net of von Neumann algebras on 
continuously parametrized bounded domains 
${\cal O}^x_s$, $s_{\min}\leq s\leq s_{\max}$, around any point $x$.  

Different than in \cite{Sal1}, it is considered 
here to be more natural that both, 
${\cal O}^x_{s_{\min}}$ and ${\cal O}^x_{s_{\max}}$
are nonempty bounded open sets, 
with their boundaries representing some kind
of horizon for observers located outside  ${\cal O}^x_{s_{\max}}$
and inside  ${\cal O}^x_{s_{\min}}$ respectively.
These bounded sets simultaneously introduce
a small and large scale regularization
naturally from the commutant duality for isotonic algebras.
In this approach we avoid to fall in an a priori conflict
like between the usual cutoffs and (general) covariance.

In Sect. 5 the commutant duality, together with the isotony property,
is used to extract the modular structure. The latter is 
related to local dilations on the net.

Sect. 6 concludes with a brief discussion of some possible
implications of the proposed structure for quantum general relativity
and a posteriori notions of time and causality.
\section{Algebraic axioms of QFT on Minkowski space} 
\setcounter{equation}{0}
Usual QFT on Minkowski space ${\cal M}$ may be formulated in terms
of a net of von Neumann algebras ${\cal R}({\cal O})$
on open sets ${\cal O}\subset {\cal M}$,
satisfying the following axioms:
\nl
{\bf Isotony}
\bear{M1}
{\cal O}_1\subset {\cal O}_2 & \imp &
{\cal R}({\cal O}_1)\subset {\cal R}({\cal O}_2)\ .
\ear
\nl
{\bf Additivity}
\bear{M2}
{\cal O} = \cup_{j} {\cal O}_j &\imp &
{\cal R}({\cal O}) = \left(\cup_{j} {\cal R}({\cal O}_j)\right)'' \ .
\ear
\nl
{\bf Causality}
\bear{M3}
{\cal O}_1\perp {\cal O}_2 &\imp&
{\cal R}({\cal O}_1)\subset {\cal R}({\cal O}_2)'\ .
\ear
\nl
{\bf Covariance}
\bear{M4}
P \ni g {\stackrel{\exists}{\mapsto}} U(g) \in U(P)&:&
{\cal R}(g{\cal O})=U(g){\cal R}({\cal O})U(g)^{-1} \ .
\ear
\nl
{\bf Spectrum Condition}
\bear{M5}
{\rm spec} U(\tau)\subset {\overline{ V^+}}
\ , &\quad&
\tau\subset P\ .
\ear
\nl
{\bf Vacuum Vector}
\bear{M6}
\exists \Omega\in{\cal H}, ||\Omega||=1 &:&
\nn\\
\mbox{\rm (cyclic)}&&
\left(\cup_{\cal O} {\cal R}({\cal O})\right) \Omega
{\stackrel{\rm dense}{\subset}} {\cal H} 
\nn\\
\mbox{\rm (invariant)}&&
U(g)\Omega=\Omega
\ , \quad
g\in P\ .
\ear
In the case of an arbitrarily curved manifold $M$,
condition \rf{M1} will be kept unmodified,
an analogue of condition \rf{M4} will be maintained,
\rf{M3} will not survive in this way, instead a new 
commutant duality will be introduced on the boundary of the net,
\rf{M5} does no longer makes sense here,
and a cyclic vector like in \rf{M6} will no longer be unique
but state dependent, 
while its gauge invariance will survive in modified form.
The present investigations will not solve the problem,
how to relate algebras on arbitrarily intersecting sets,
whence \rf{M2} will be just ignored here.
\section{Covariant nets of algebras} 
\setcounter{equation}{0}
Given a differentiable manifold $M$,
a collection $\{{\cal A}({\cal O})\}_{{\cal O}\in M}$ 
of $*$-algebras  ${\cal A}({\cal O})$
on bounded open sets ${\cal O} \in M$ 
is called a {\em net of $*$-algebras},
iff
\beq{iso}
{\cal O}_1\subset {\cal O}_2\imp
{\cal A}({\cal O}_1)\subset {\cal A}({\cal O}_2)\ .
\eeq
The net is sometimes also denoted by ${\cal A}^{}_{}:=
{\bigcup_{\cal O}} {\cal A}^{}_{}({\cal O})$.
Selfadjoint elements of ${\cal A}({\cal O})$ 
may be interpreted as possible measurements in $\cal O$. 

A net of algebras on $M$ is $\Diff(M)$ covariant, if it reflects
the $\Diff(M)$ covariance  
of the underlying manifold $M$.
$\Diff(M)$ then acts by algebraic isomorphisms on 
${\cal A}:={\bigcup_{\cal O}} {\cal A}({\cal O})$, 
i.e. each diffeomorphism
$\chi\in \Diff(M)$ induces an algebraic isomorphism 
$\alpha_{\chi}$ such that
\begin{equation}
\label{I1}  
\alpha_{\chi}({\cal A}({\cal O}))={\cal A}(\chi({\cal O}))\ .
\end{equation}
Two sets ${\cal O}_1$ and ${\cal O}_2$, related
by a topological isomorphism (e.g. a diffeomorphism) $\chi$
such that $\chi({\cal O}_1)={\cal O}_2$,
may be identified straightforwardly only if there are
no further obstructing relations between them. A relation 
like ${\cal O}_1\subset{\cal O}_2$, in addition to the previous one, 
implies that 
${\cal O}_1$ and ${\cal O}_2$ have to be considered as 
topologically isomorphic, though non-identical, sets.
A similar situation holds on the level of algebras.
Isotony \rf{iso} in connection with covariance \rf{I1}
implies that ${\cal A}({\cal O}_1)$ and ${\cal A}({\cal O}_2)$
are isomorphic but non-identical algebras.
Therefore it was a misleading abuse of terminology
in previous papers\cite{Ra1,RaSa}
to call $\alpha_{\chi}$ an algebraic automorphism 
(as e.g. \cite{Ha}) although
the situation is more complicated in general.
In the following, algebras related 
simultaneously by isotonic inclusion
and an algebraic isomorphism  are, more correctly, called
just isomorphic rather than automorphic algebras.

The state of a physical system is mathematical described by a 
positive linear functional  ${\omega}$ on $\cal A$. 
Given the state ${\omega}$, one
gets via the GNS construction a representation $\pi^{\omega}$ of $\cal A$ 
by a net of operator algebras on a Hilbert space ${\cal H}^{\omega}$ with 
a cyclic vector $\Omega^{\omega}\in {\cal H}^{\omega}$. The
GNS representation $(\pi^{\omega}, {\cal H}^{\omega}, \Omega^{\omega})$ 
of any state $\omega$ has an associated folium ${\cal F}^{\omega}$, 
given as the family of those states $\omega_\rho:=\tr\rho\pi^{\omega}$ 
which are defined by positive trace class
operators $\rho$ on ${\cal H}^{\omega}$. 

Once a physical state $\omega$ 
(which implicitly contains all peculiarities of a particular
observation)
has been specified, 
one can consider in each
algebra ${\cal A}({\cal O})$ the equivalence relation
\begin{equation}
A\sim B \ \ :\Iff \ \ 
{\omega}^{\prime}(A-B)=0,\ \ \forall  {\omega}^{\prime}\in{\cal F}^{\omega}\ .
\end{equation}
These equivalence relations generate a two-sided ideal 
\beq{ideal}
{\cal I}^{\omega}({\cal O}):=\{A\in{\cal A}({\cal O})\vert
{\omega}^{\prime}(A)=0\}
\eeq
in ${\cal A}({\cal O})$.
The (dynamically relevant) state dependent algebra of { observables} 
${\cal A}^{\omega}_{}({\cal O}):=\pi^{\omega}({\cal A}({\cal O}))$ may
be constructed from
the (kinematically relevant) algebra of 
observation procedures ${\cal A}({\cal O})$ by taking 
the quotient
\begin{equation}
{\cal A}^{\omega}_{}({\cal O}) =
{\cal A}({\cal O})/{\cal I}^{\omega}({\cal O})\ .
\end{equation}
The net of state-dependent algebras then is also denoted as
${\cal A}^{\omega}_{}:=
{\bigcup_{\cal O}} {\cal A}^{\omega}_{}({\cal O})$.
By construction, any diffeomorphism $\chi\in \Diff(M)$ induces an 
algebraic isomorphism $\alpha_{\chi}$ of the observation procedures.
Nevertheless, for a given state $\omega$,  
the action of $\alpha_{\chi}$ will 
in general {\em not} leave ${\cal A}^{\omega}_{}$
invariant.
In order to satisfy
\begin{equation}
\label{D1}
\alpha_{\chi}({\cal A}^{\omega}_{}({\cal O}))=
{\cal A}^{\omega}_{}(\chi({\cal O}))\ .
\end{equation}
the ideal
${\cal I}^{\omega}({\cal O})$ must transform covariantly, i.e.
the diffeomorphism ${\chi}$ must satisfy
the condition
\begin{equation}
\label{D2}
\alpha_{\chi}({\cal I}^{\omega}({\cal O}))=
{\cal I}^{\omega}(\chi({\cal O}))\ 
\end{equation}
for some algebraic isomorphism $\alpha_{\chi}$.
Due to non-trivial constraints \rf{D2}, the (dynamical) algebra of 
observables, constructed with respect to
the folium ${\cal F}^\omega$, does in general no longer exhibit 
the full $\Diff(M)$ symmetry of the (kinematical) observation procedures. 
The symmetry of the observables is dependent on (the folium of) the state
$\omega$. 
Therefore, the selection of a folium of states
${\cal F}^\omega$, induced by the actual choice of a state $\omega$,
results immediately in a breaking of the $\Diff(M)$ symmetry.
The diffeomorphisms which satisfy 
the constraint condition (\ref{D2}) form a subgroup.  
This effective symmetry group is called  
the {\em dynamical group} of the state $\omega$.
$\alpha_{\chi}$ is called a {\em dynamical} isomorphism
(w.r.t. the given state $\omega$) w.r.t. $\chi$, if (\ref{D2})
is satisfied.

The remaining dynamical symmetry group, depending
on the folium ${\cal F}^\omega$ of states related to $\omega$,
has two main aspects which we have to examine
in order to specify the physically admissible states:
Firstly, it is necessary to specify its state dependent  
algebraic action on the net of observables. Secondly, one has to
find a geometric interpretation for the group and its action on $M$.

If we consider the dynamical group as an {\em inertial}, 
and therefore global, manifestation of dynamically ascertainable 
properties of observables, 
then its (local) action should be correlated with (global) 
operations on the whole net of observables.
This implies that at least some of the dynamical
isomorphisms $\alpha_\chi$ are not inner.
(For the case of causal nets of algebras 
it was actually already shown 
that, under some additional assumptions,
the isomorphisms of the algebras are in general not inner\cite{Wo}.)

Note that one might consider instead of the net of observables
${\cal A}^{\omega}_{}({\cal O})$  
the net of associated von Neumann algebras
${\cal R}^{\omega}_{}({\cal O})$, which can be defined even for
unbounded ${\cal A}^{\omega}_{}({\cal O})$, if we take 
from the modulus of the von Neumann closure  
$({{\cal A}^{\omega}_{}}({\cal O}))''$ 
all its spectral projections\cite{FrHa}.
Then the isotony (\ref{iso}) induces a likewise isotony of the net 
${\cal R}^{\omega}_{}:=
{\bigcup_{\cal O}} {\cal R}^{\omega}_{}({\cal O})$
of von Neumann algebras.

\section{Local dilations} 
\setcounter{equation}{0}
In the following I want to exhibit a possibility to introduce
simultaneously small and large scale cutoff regularizations
on the net of von Neumann algebras.
This essentially exploits a local partial ordering on the net,
which is induced by the isotony property.

Let us now make use of the given ($C^\infty$) topological structure of
$M$ and choose at a given point $x\in M$ a topological basis
of nonzero open sets 
${\cal O}^{x}_s\ni x$,
parametrized by a real parameter $s$ with $0<s<\infty$, 
such that
\begin{equation}\label{strictin}
s_1<s_2 \quad \Iff \quad \cl ({\cal O}^{x}_{s_1}) 
{\subset} {\cal O}^{x}_{s_2} 
\label{inc}
\end{equation}
and 
\begin{equation}
s\to 0 \quad \Iff \quad \cl ({\cal O}^{x}_{s}) {\to} \{x\} \ .
\end{equation}
The standard inclusion 
${\cal O}^{x}_{s_1}{\subset} {\cal O}^{x}_{s_2}$ 
(used  previously\cite{Ra1})
does not exclude the possibility that
$\partl{\cal O}^{x}_{s_1} {\cap} \partl{\cal O}^{x}_{s_2}\neq \emptyset$.
Since, for the following, this would be slightly pathological, 
condition \rf{strictin} uses here
$\cl ({\cal O}^{x}_{s_1})
{\subset} {\cal O}^{x}_{s_2}$ 
as a slightly more strict inclusion instead.
 
Let the parameter $s$ be restricted by
$0<s_{\min,x}<s<s_{\max,x}<\infty$.
Exploiting local reparametrization invariance,
one may assume 
\bear{univers}
s_{\min,x}=s_{\min} ,\quad s_{\max,x}=s_{\max} \qquad \forall x\in M \ ,   
\ear
without loss of generality.
Then, for each $x\in M$, open sets ${\cal O}^{x}_{s}$ with 
$s\in ]s_{\min},s_{\max}[$ generate local cobordisms between
$\partial {\cal O}^{x}_{s_{\min}}$
and $\partial {\cal O}^{x}_{s_{\max}}$, and
the isotony property \rf{iso} 
implies that 
\bear{Rinc}
&s_{\min}<s_1<s_2<s_{\max} \imp&
\nn\\
&{\cal R}^{\omega}_{}({\cal O}^{x}_{s_{\min}})
{\ \subset\ } {\cal R}^{\omega}_{}({\cal O}^{x}_{s_1})
{\ \subset\ } {\cal R}^{\omega}_{}({\cal O}^{x}_{s_2})
{\ \subset\ } {\cal R}^{\omega}_{}({\cal O}^{x}_{s_{\max}}) &\ .
\ear
Here, any diffeomorphism ${\cal O}^{x}_{s_1}\mapsto{\cal O}^{x}_{s_2}$
is a {\em local dilation} 
at $x\in M$ from ${\cal O}^{x}_{s_1}$ to ${\cal O}^{x}_{s_2}$.
Note that  
all local dilations 
which preserve covariance of \rf{Rinc}
must leave invariant ${\cal O}^{x}_{\min}$
and ${\cal O}^{x}_{\max}$. 

Now, a commutant duality relation between
the inductive limits given by the minimal and maximal algebras
is introduced,
\begin{equation}
\label{dmin}
{\cal R}^{\omega}_{}({\cal O}^{x}_{s_{\min}})
=\left({{\cal R}^{\omega}_{}({\cal O}^{x}_{s_{\max}})}\right)'\ ,
\end{equation}
where ${\cal R}'$ denotes the commutant of ${\cal R}$ within some
${\cal R}_{\max}\supset {\cal R}$. 
Then the bicommutant theorem (${\cal R}''={\cal R}$) 
implies that likewise also
\begin{equation}
\label{dmax}
{\cal R}^{\omega}_{}({\cal O}^{x}_{s_{\max}})
=\left({{\cal R}^{\omega}_{}({\cal O}^{x}_{s_{\min}})}\right)'\ .
\end{equation}
If one now demands that all maximal (or all minimal) algebras
are isomorphic to each other, independently of the choice of $x$ and the
open set ${\cal O}^{x}_{s_{\max}}$ 
(resp. ${\cal O}^{x}_{s_{\min}}$),
then by (\ref{dmin}) (resp. (\ref{dmax})) also all minimal (resp. maximal)
algebras are isomorphic to each other.  
The isomorphism class is then an abstract universal minimal resp. maximal 
algebra, denoted by 
${\cal R}^{\omega}_{{\min}}$ and ${\cal R}^{\omega}_{{\max}}$ 
respectively. 

If, like in the following, the commutant is always  
taken within ${\cal R}^{\omega}_{{\max}}$.
the duality (\ref{dmin}) implies that 
${\cal R}^{\omega}_{{\min}}$ is Abelian.
(Note however, that one should also keep in mind the possibility
to take the commutant w.r.t. to some larger algebra
 ${\cal R}^{\omega}_B\supset {\cal R}^{\omega}_{{\max}}$.
Such a choice would possibly include 
further correlations outside the observable range;
it is not considered further here.)

By isotony and (\ref{inc}) together with \rf{univers}, 
the mere existence of 
${\cal R}^{\omega}_{\min}$ 
resp. ${\cal R}^{\omega}_{\max}$ implies
the existence of nontrivial sets 
${\cal O}^{x}_{s_{\min}}$ resp. ${\cal O}^{x}_{s_{\max}}$
at any  $x\in M$.
By \rf{univers} we already gauged the size of all these sets
to $s_{\min}$ resp. $s_{\max}$, i.e.
to a common size (as measured by the parameter $s$) of
independently of $x\in M$.
So in this case $s_{\min}$ and $s_{\max}$ really denote 
an universal small resp. large scale cutoff.
Note that, in the context of Sect. 3, the universality assumption 
\rf{univers} is indeed nontrivial, because local diffeomorphisms 
consistent 
with the structure above must preserve $s_{\min}$,  $s_{\max}$,
and the monotony of the ordered set $]s_{\min},s_{\max}[$.
The number $s\in]s_{\min},s_{\max}[$
parametrizes the partial order of the net of algebras
spanned between the inductive limits 
${\cal R}^{\omega}_{\min}$ and ${\cal R}^{\omega}_{\max}$.
(Strictly speaking, these inductive limits are not part of the 
diffeorphism invariant net itself !)

Although in local QFT usually the support of an algebra and
that of its commutant are not at all related,
it might be nevertheless instructive to consider the
case where (sufficiently large) algebras of the net satisfy 
\begin{equation}
\label{cominc}
\left({{\cal R}^{\omega}_{}({\cal O}_s^{x})}\right)'
\subset {\cal R}^{\omega}_{}({\cal O}_s^{x})\ .
\end{equation}
Then, with the center of ${\cal R}^{\omega}_{}({\cal O}_s^{x})$
defined as 
${\cal Z}\left({\cal R}^{\omega}_{}({\cal O}_s^{x})\right):=
{\cal R}^{\omega}_{}({\cal O}_s^{x})\cap
\left({{\cal R}^{\omega}_{}({\cal O}_s^{x})}\right)'$,
one obtains 
${\cal Z}\left({\cal R}^{\omega}_{}({\cal O}_s^{x})\right)=
\left({{\cal R}^{\omega}_{}({\cal O}_s^{x})}\right)'=
{\cal Z}\left(({{\cal R}^{\omega}_{}({\cal O}_s^{x})})'\right)$,
and especially
${\cal Z}\left({\cal R}^{\omega}_{\max}\right)=
{{\cal R}^{\omega}_{\min}}=
{\cal Z}\left({\cal R}^{\omega}_{\min}\right)$.
So, for a pair of commutant dual algebras satisfying
Eq. (\ref{cominc}), the smaller one is always
Abelian, namely it is the center of the bigger one.
With (\ref{cominc}), the isotony of the net implies the existence 
of an algebra ${\cal Z}^{\omega}$ which is {\em maximal Abelian}, 
in other words
commutant selfdual, satisfying ${\cal Z}^{\omega}=({\cal Z}^{\omega})'=
{\cal Z}({\cal Z}^{\omega})$. This algebra is given explicitly via the
Abelian net of all centers, ${\cal Z}^{\omega}
:={\bigcup_{\cal O}}{\cal Z}\left({\cal R}^{\omega}_{}({\cal O})\right)$.
${\cal Z}^{\omega}$, located on an underlying set ${\cal O}^{x}_{s_{z}}$
of intermediate size s.th. $s_{\min}<s_z<s_{\max}$,  separates
the small Abelian algebras 
${{\cal R}^{\omega}_{}({\cal O}_s^{x})}=
{\cal Z}\left({\cal R}^{\omega}_{}({\cal O}_s^{x})\right)$,
with $s\leq s_z$, from larger
non-Abelian algebras 
${{\cal R}^{\omega}_{}({\cal O}_s^{x})}=
\left({\cal Z}({\cal R}^{\omega}_{}({\cal O}_s^{x}))\right)'$,
with $s>s_z$. 

For a net subject to (\ref{cominc}), its lower end is Abelian, 
whence observations on small regions  with $s\leq s_z$
are expected to be rather classical.
Nevertheless, for increasing size  $s>s_z$, 
there might well exist a non-trivial
quantum (field) theory 
(in fact it was shown\cite{Wo} that, for  causal nets, the
algebras of QFT are not Abelian and not finite-dimensional).
For quantum general relativity there might indeed be a 
kinetical substructure\cite{Sal2}, with classical elementary constituents 
spanning an Abelian algebra.
It is interesting in this context that the Abelian algebra 
of free loops in quantum general relativity
provides indeed such classical constituents\cite{Ro,AshIsh}.

Nevertheless, the following investigations all hold independently
from relation \rf{cominc}. Indeed we will see below, that
 \rf{cominc} can only make sense if we take the commutant 
w.r.t. some algebra essentially larger than ${\cal R}^{\omega}_{\max}$.
\section{Modular structure and dilations} 
\setcounter{equation}{0}
If we consider the small and large scale cutoffs as introduced above,
it should be clear that only regions 
of size $s\in ]s_{\min},s_{\max}[$ are admissible for measurement.  
The commutant duality between 
${\cal R}^{\omega}_{\min}$ and ${\cal R}^{\omega}_{\max}$
inevitably yields large scale correlations 
in the structure of any physical state $\omega$ on
any admissible region ${\cal O}^{x}_{s}$ of measurement at $x$.
Let us assume here that $\omega$ is properly correlated,
i.e. the GNS vector $\Omega^\omega$ is already cyclic under 
${\cal R}^{\omega}_{\min}$. 
Then, by duality,  it is separating for 
${\cal R}^{\omega}_{\max}={{\cal R}^{\omega}_{\min}}'$.
Furthermore $\Omega^\omega$  is also cyclic under  
${\cal R}^{\omega}_{\max}$, and hence
separating for ${\cal R}^{\omega}_{\min}$.

So $\Omega^{\omega}$ is a cyclic and separating vector for 
${\cal R}^{\omega}_{\min}$ and ${\cal R}^{\omega}_{\max}$, 
and by isotony also for any
local von Neumann algebra ${\cal R}^{\omega}_{}({\cal O}^{x}_{s})$.

As a further consequence, on any region ${\cal O}^{x}_{s}$, 
the Tomita operator $S$ and  its conjugate $F$ 
can be defined densely by
\begin{equation}
\label{T0}
S A \Omega^{\omega}:= A^{*} \Omega^{\omega} \ \ \mbox{for}\ \ A\in 
{\cal R}^{\omega}_{}({\cal O}^{x}_{s})\ ,
\end{equation}
\begin{equation}
\label{T1}
F B \Omega^{\omega}:=B^{*} \Omega^{\omega}
 \ \ \mbox{for}\ \ B\in 
{{\cal R}^{\omega}_{}({\cal O}^{x}_{s})}' \ . 
\end{equation}
The  closed Tomita operator $S$ has a polar decomposition 
\begin{equation}
S=J\Delta^{1/2} \ ,
\label{T2}
\end{equation}
where  $J$ is antiunitary and $\Delta:=FS$ is the self-adjoint, positive 
modular operator.
The Tomita-Takesaki theorem \cite{Ha} provides us with a one-parameter 
group of state dependent isomorphisms $\alpha^{\omega}_t$ on 
${\cal R}^{\omega}_{}({\cal O}^{x}_{s})$,
defined by
\begin{equation}
\alpha^{\omega}_t (A)= \Delta^{-it}\ A\ \Delta^{it}\ ,   \ \ \mbox{for}\ \ 
A\in{\cal R}^{\omega}_{\max} \ .
\label{T3}
\end{equation}
So, as a consequence of commutant duality and isotony assumed above,  
we obtain here a strongly continuous unitary 
implementation  of the modular group of $\omega$,
which is defined by  the $1$-parameter family of isomorphisms (\ref{T3}),
given as conjugate action of operators
$e^{-it\ln\Delta}$, ${t\in\R}$.
By (\ref{T3}) the modular group, for a state $\omega$  
on the net of von Neumann algebras, 
defined by ${\cal R}^{\omega}_{\max}$,  
might be considered as a $1$-parameter subgroup of the dynamical group.
Note that, with Eq. (\ref{T1}), in general,
the modular operator $\Delta$ is not located on
${\cal O}^{x}_{s}$. Therefore,  in general, the modular isomorphisms 
(\ref{T3}) are not inner.
The modular isomorphisms are known to act as 
{inner} isomorphisms, iff the von Neumann algebra 
${\cal R}^{\omega}_{}({\cal O}^{x}_{s})$
generated by $\omega$
contains only semifinite factors (type I and II), i.e. 
$\omega$ is a semifinite trace.

Above we considered concrete von Neumann algebras
${\cal R}^{\omega}_{}({\cal O}^{x}_{s})$, which are in fact 
operator representations of an abstract von Neumann algebra
${\cal R}$ on a GNS Hilbert space ${\cal H}^{\omega}$ w.r.t.
a faithful normal state $\omega$.
In general, different faithful normal states generate different 
concrete von Neumann algebras and different modular isomorphism groups
of the same abstract  von Neumann algebra.

The outer modular isomorphisms form the cohomology group  
$\Out {\cal R}:=\Aut {\cal R}/\Inn {\cal R}$
of modular isomorphisms modulo inner modular isomorphisms. 
This group is characteristic
for the types of factors contained in the von Neumann algebra\cite{Co}.
Per definition $\Out {\cal R}$ is trivial for inner isomorphisms.
Factors of type III${}_1$ yield $\Out {\cal R}=\R$.

In the case of thermal equilibrium states, 
corresponding to factors 
of type III${}_1$, there is a distinguished
$1$-parameter group of outer modular isomorphisms,
which is a subgroup of the dynamical group. 

Looking for a geometric interpretation for this subgroup, 
parametrized by $\R$, it should not be a coincidence
that our partial order defined above is parametrized
by open intervals
(namely $]s_{\min},s_{\max}[$ for the full net
and, in the case of \rf{cominc}, $]s_{z},s_{\max}[$ 
for the non-Abelian part), 
and hence diffeomorphically likewise by $\R$.  
This way, dilations  of the open sets ${\cal O}^{x}_{s}$
within the open interval
may give a geometrical meaning to the $1$-parameter group of outer modular 
isomorphisms of thermal equilibrium states. 
Even more, this gives support to the hypothesis\cite{CoRo} of a thermal time.     
Indeed, a local equilibrium state might be characterized
as a KMS state\cite{Ha,BrRo} over the algebra of observables 
on a (suitably defined) double cone, whence the $1$-parameter modular group 
in the KMS condition might be related to the time evolution.
Note that, for double cones, a partial order
can be related to the split property of the algebras\cite{Ba2}.

Now, the details of the isotony condition \rf{iso}, in relation
to the modular invariance \rf{T3},
allow rather immediately to draw some further conclusions,
which have not yet been spelled out in previous 
investigations\cite{Ra1,RaSa}.
First assume strict isotony, i.e.
\begin{equation}
s_1<s_2 \imp
{\cal R}^{\omega}_{}({\cal O}^{x}_{s_1})
{\ {{\subset\vspace*{-0.0mm}}\atop{\vspace*{0.0mm}\neq}}\ } 
{\cal R}^{\omega}_{}({\cal O}^{x}_{s_2}) \ .
\end{equation}
Covariance w.r.t. local dilations then implies
isomorphic algebras
${\cal R}^{\omega}_{}({\cal O}^{x}_{s_1})\iso
{\cal R}^{\omega}_{}({\cal O}^{x}_{s_2})$, for $s_{\min}<s_1<s_2<s_{\max}$,
whence, in particular, all algebras have the same von Neumann type.
Obviously, here the condition \rf{cominc} would
lead to a totally Abelian net, if the commutant is not taken in 
a larger algebra  ${\cal R}^{\omega}_B\supset {\cal R}^{\omega}_{{\max}}$.
Therefore \rf{cominc} is not further considered here.
With the commutant duality \rf{dmin} w.r.t. ${\cal R}^{\omega}_{\max}$ above, 
${\cal R}^{\omega}_{\min}$
is Abelian. Hence, the net contains non-Abelian algebras (in particular
those with type III${}_1$) only if minimal and maximal sets and algebras 
are excluded from the net.
${\cal R}^{\omega}_{\min}$ and ${\cal R}^{\omega}_{\max}$
then only exist as inductive limits of a net of isomorphic algebras,
while $\partl{\cal O}^{x}_{\min}$ resp. $\partl{\cal O}^{x}_{\max}$   
then are horizon-like boundaries of the open manifold supporting the net.
Here, only local regions with 
$s\in ]s_{\min},s_{\max}[$ are admissible for measurement.

Note that in the case where the manifold carries 
a (pseudo-)metric $g$, the net must not only be
consistent with local dilations of open sets, but also with 
local dilations of the metric, which take the form of
conformal transformations $g(x)\mapsto e^{2\phi(x)}g(x)$,
with smooth scale field $\phi$ on $M$.
If g is consistent with $M=]s_{\min},s_{\max}[\times\Sigma$
(e.g. by global hyperbolicity),
consistency with covariance demands that $\phi$ is homogeneous on the
"horizons", i.e. 
$\phi(s_{\min},y)\equiv \phi(s_{\min})$
and
$\phi(s_{\max},y)\equiv \phi(s_{\max})$
for all $y\in\Sigma$.
\section{Discussion} 
\setcounter{equation}{0}
It is  clear now that (\ref{cominc})
does not make sense, if the commutant for all algebras of the net 
is taken w.r.t. ${\cal R}^{\omega}_{\max}$, whence all algebras
would be Abelian. Even if we drop  (\ref{cominc}),
with this choice of commutant, 
at least ${\cal R}^{\omega}_{\min}$ is Abelian.

In principle, there remains the possibility,
to take the commutant w.r.t. a larger algebra  
${\cal R}^{\omega}_B\supset {\cal R}^{\omega}_{{\max}}$.
Then we may obtain also non-Abelian ${\cal R}^{\omega}_{\min}$.
However, further examinations are necessary until we can decide clearly
which way is the better. 

Note however that only the commutant duality
(\ref{dmin}), but not  (\ref{cominc}), was essential for
the extraction of the modular group.
If we assume von Neumann algebras (with factors) of type
III${}_1$,  or likewise the
existence of local equilibrium states,  
with the the partial order and related dilations above,
it is possible to make natural choices for time and causality. 

A thermal time may be obtained from
the geometric notion of dilations of the open sets. 
For any $x\in M$, the parameter $s$ measures the extension
of the sets ${\cal O}^{x}_{s}$. 
As accessibility regions for a local 
measurement in $M$, these sets naturally increase with time. 
Hence it is natural to suggest 
that the parameter $s$ might be
related to a (thermal) time $t$
such that, for any set ${\cal O}^{x}_{s}$,
$s>s_{\min}$, we have $t<s$ within the set and 
$t=s$ on the boundary $\partial{\cal O}^{x}_{s}$. 

This thermal time here is related to growing non-identical algebras
of increasing support.
If covariance is kept as a condition, these algebras are 
nevertheless isomorphic.
A non-isomorphic growth of algebras requires to release
the covariance condition.
Then, the growth of abstract (isomorphism classes) of algebras
may define an arrow of time. 

For the ultralocal case 
($s_{\min}\to 0$, without UV cutoff), 
it is possible to construct the causal structure for a space-time from 
the corresponding net of operator algebras\cite{Keyl}.
Let us consider here the (a priori given) underlying manifold $M$ of the net.
Locally around any point $x\in M$ one may induce  open 
double cones  as the 
pullback of the standard double cone which is 
the conformal model of  Minkowski space.
These open double cones then carry natural
notions of time and causality, 
which are preserved under dilations.  
Therefore it seems natural to introduce locally around any $x\in M$ 
a causal structure and time by specializing the open sets to be
open double cones ${\cal K}^{x}_{s}$ located at $x$, with
time-like extension $2s$ between the ultimate past event $p$ and
the ultimate future event $q$ involved in any measurement in 
${\cal K}^{x}_{s}$ at $x$ 
(time $s$ between $p$ and $x$, and likewise between $x$ and $q$). 
Since the open double cones form a basis for the local topology of $M$, 
we might indeed consider equivalently the net of algebras located
on open sets 
\beq{dc}
{\cal O}^{x}_{s}:={\cal K}^{x}_{s}\ .
\eeq
Although some (moderate form of) locality might be indeed an 
indispensable principle within any reasonable theory of observations,
it is nevertheless an important but difficult question, 
under which consistency conditions
a local notion of time and causality might be extended 
from nonzero local environments of individual points to global regions.
This is of course also related to the non-trivial open question,
how open neighbourhoods of different points $x_1\neq x_2$
should be related consistently. Although, in a radical attempt to avoid some
part of these difficulties, one might try to replace the notion of points
and local regions by more abstract algebraic concepts,  
a final answer to these questions has not yet been found.
At least it seems natural that,
on manifolds with no 
causal relations (like pure space without time), the net should satisfy  
a disjoint compatibility condition,
\begin{equation}\label{discom}
{\cal O}_1\cap {\cal O}_2=\emptyset
\quad \imp \quad [{\cal A}({\cal O}_1), {\cal A}({\cal O}_2)]=0 .
\end{equation}
This condition is e.g. also 
satisfied for Borchers algebras. 
Of course the inverse of (\ref{discom})
is not true in general.

\section*{Acknowledgments}
The participation at this conference was kindly supported by 
the Deutsche Forschungsgemeinschaft (DFG).

\np

\end{document}